\documentclass[conference]{IEEEtran}
\IEEEoverridecommandlockouts
\usepackage{cite}
\usepackage{amsmath,amssymb,amsfonts}
\usepackage{algorithmic}
\usepackage{graphicx}
\usepackage{textcomp}
\usepackage{xcolor}
\usepackage{subcaption}
\def\BibTeX{{\rm B\kern-.05em{\sc i\kern-.025em b}\kern-.08em
    T\kern-.1667em\lower.7ex\hbox{E}\kern-.125emX}}
\begin{document}

\title{Conditioning Autoencoder Latent Spaces for Real-Time Timbre Interpolation and Synthesis
}

\author{\IEEEauthorblockN{Joseph T Colonel}
\IEEEauthorblockA{\textit{Queen Mary University of London} \\
London, UK \\
j.t.colonel@qmul.ac.uk}
\and
\IEEEauthorblockN{Sam Keene}
\IEEEauthorblockA{\textit{Cooper Union}} 
New York, USA \\
keene@cooper.edu}

\maketitle

\begin{abstract}
We compare standard autoencoder topologies' performances for timbre generation. We demonstrate how different activation functions used in the autoencoder's bottleneck distributes a training corpus's embedding. We show that the choice of sigmoid activation in the bottleneck produces a more bounded and uniformly distributed embedding than a leaky rectified linear unit activation. We propose a one-hot encoded chroma feature vector for use in both input augmentation and latent space conditioning. We measure the performance of these networks, and characterize the latent embeddings that arise from the use of this chroma conditioning vector. An open source, real-time timbre synthesis algorithm in Python is outlined and shared.  
\end{abstract}


\section{Introduction}
Timbre refers to the perceptual qualities of a musical sound distinct from its amplitude and pitch. It is timbre that allows a listener to distinguish between a guitar and a violin both producing a concert C note.  Moreover, a musician's ability to create, control, and exploit new timbres has led, in part, to the surge in popularity of pop, electronic, and hip hop music.  

New algorithms for timbre generation and sound synthesis have accompanied the rise to prominence of artificial neural networks. GANSynth \cite{gansynth}, trained on the NSynth dataset \cite{nsynth2017}, uses generative adversarial networks to output high-fidelity, locally coherent outputs. Furthermore, GANSynth's global latent conditioning allows for interpolations between two timbres. Other works have found success in using Variational Autoencoders (VAEs) \cite{universal} \cite{bijective} \cite{assisted}, which combine autoencoders and probabilistic inference to generate new audio. Most recently, differential digital signal processing has shown promise by casting common modules used in signal processing into a differentiable form \cite{ddsp}, where they can be trained with neural networks using stochastic optimization. 

The complexity of these models require powerful computing hardware to train, hardware often out of reach for musicians and creatives. When designing neural networks for creative purposes one must strike a three-way balance between the expressivity of the system, the freedom given to a user to train and interface with the network, and the computational overhead needed for sound synthesis. One successful example we point to in the field of music composition is MidiMe \cite{midime}, which allows a composer to train a VAE with their own scores on a subspace of a larger, more powerful model. Moreover, these training computations take place on the end user's browser.  

Our previous work has tried to strike this three-way balance as well \cite{colonel} \cite{colonel2}, by utilizing feed-forward neural network autoencoder architectures trained on Short-Time Fourier Transform (STFT) magnitude frames. This work demonstrated how choice of activation functions, corpora, and augmentations to the autoencoder's input could improve performance for timbre generation. However, we found upon testing that the autoencoder's latent space proved difficult to control and characterize. Also, we found that our use of a five-octave MicroKORG corpus encouraged the autoencoder to produce high-pitched, often uncomfortable tones.

This paper introduces a chroma-based input augmentation and skip connection to help improve our autoencoder's reconstruction performance with little additional training time. A one-octave MicroKORG corpus as well as a violin-based corpus are used to train and compare various architectural tweaks. Moreover, we show how this skip connection conditions the autoencoder's latent space so that a musician can shape a timbre around a desired note class. A full characterization of the autoencoder's latent space is provided by sampling from meshes that span the latent space. Finally, a real-time, responsive implementation of this architecture is outlined and made available in Python. 

\section{Autoencoding Neural Networks}

An autoencoding neural network (i.e. autoencoder) is a machine learning algorithm that is typically used for unsupervised learning of an encoding scheme for a given input domain, and is comprised of an encoder and a decoder \cite{stacked}. For the purposes of this work, the encoder is forced to shrink the dimension of an input into a latent space using a discrete number of values, or ``neurons.'' The decoder then expands the dimension of the latent space to that of the input, in a manner that reconstructs the original input. 

In a single layer model, the encoder maps an input vector $x \in \mathbb{R}^d $ to the hidden layer $y \in \mathbb{R}^e $, where $d>e$. Then, the decoder maps $y$ to $\hat{x} \in \mathbb{R}^d $. In
this formulation, the encoder maps $x \rightarrow y $ via
\begin{equation} \label{Wx:encode}
y = f(Wx+b)
\end{equation}
where $W \in \mathbb{R}^{(e \times d)}$, $b \in \mathbb{R} ^e $, and
$f(\cdotp)$ is an activation function that imposes a non-linearity in
the neural network. The decoder has a similar formulation:
\begin{equation} \label{Wx:decode}
\hat{x} = f(W_{\text{out}} y+b_{\text{out}})
\end{equation}
with $W_{\text{out}} \in \mathbb{R} ^{(d \times e)}$, $b_{out} \in \mathbb{R} ^d $. 

A multi-layer autoencoder acts in much the same way as a single-layer autoencoder. The encoder contains $n>1$ layers and the decoder contains $m>1$ layers. Using Equation \ref{Wx:encode} for each mapping, the encoder maps $x \rightarrow x_{1} \rightarrow \ldots  \rightarrow x_{n} $. Treating $x_{n}$ as $y$ in Equation \ref{Wx:decode}, the decoder maps $x_{n} \rightarrow x_{n+1} \rightarrow \ldots \rightarrow x_{n+m} = \hat{x}$. 

The autoencoder trains the weights of the $W$'s and $b$'s to minimize some cost function. This cost function should minimize the distance between input and output values. The choice of activation functions $f(\cdotp)$ and cost functions depends on the domain of a given task.

\section{Network Design and Topology}

We build off of previous work to present our current network architecture.

\subsection{Activations}
The sigmoid function

\begin{equation}
f(x) = \frac{1}{1+ e^{-x}}
\end{equation}

and rectified linear unit (ReLU)

\begin{equation}
f(x) = 
\left\{
\begin{array}{ll}
      0 &,  x < 0 \\
      x &,  x \geq 0 \\
\end{array} 
\right. \end{equation}

are often used to impose the nonlinearities $f(\cdotp)$ in an autoencoding neural network. A hybrid autoencoder topology cobmining both sigmoid and ReLU activations was shown to outperform all-sigmoid and all-ReLU models in a timbre encoding task \cite{colonel}. However, this hybrid model often would not converge for a deeper autoencoder \cite{colonel2}. 

More recently, the leaky rectified linear unit (LReLU) \cite{LReLU}

\begin{equation}
f(x) = 
\left\{
\begin{array}{ll}
      \alpha x &,  x < 0 \\
      x &,  x \geq 0 \\
\end{array} 
\right. \end{equation}

has been shown to avoid both the vanishing gradient problem introduced by using the sigmoid activation \cite{vangrad} and the dying neuron problem introduced by using the ReLU activation \cite{dying}. The hyperparameter $\alpha$ is typically small, and in this work fixed at $0.1$.

\subsection{Chroma Based Input Augmentation}
The work presented in \cite{colonel2} showed how appending descriptive features to the input of an autoencoder can improve reconstruction performance, at the cost of increased training time. More specifically, appending the first-order difference of the training example to the input was shown to give the best reconstruction performance, at the cost of doubling the training time. Here, we suggest a basic chroma-based feature augmentation to help the autoencoder. 

Chroma-based features capture the harmonic information of an input sound by projecting the input's frequency content onto a set of chroma values \cite{chroma}. Assuming a twelve-interval equal temperment Western music scale, these chroma values form the set \{C, C\#, D, D\#, E, F, F\#, G, G\#, A, A\#, B\}. A chromagram can be calculated by decomposing an input sound into 88 frequency bands corresponding to the musical notes A0 to C8. Summing the short-time mean-square power across $N$ frames for each sub-band across each note (i.e. A0-A7) yields a $12 \times N$ chromagram. 

In this work, a one-hot encoded chroma representation is calculated for each training example by taking its chromagram, setting the maximum chroma value to $1$, and setting every other chroma value to $0$. While this reduces to note conditioning in the case of single-note audio, this generalizes to the dominant frequency of a chord or polyphonic mixture. Furthermore this feature can be calculated on an arbitrary corpus, which eliminates the tedious process of annotating by hand.

\subsection{Hidden Layers and Bottleneck Skip Connection}
This work uses a slight modification of the geometrically decreasing/increasing autoencoder topology proposed in \cite{colonel2}. All layers aside from the bottleneck and output layers use the LReLU activation function. The output layer uses the ReLU, as all training examples in the corpus take strictly non-negative values. For the bottleneck layer, models are trained separately with all-LReLU and all-sigmoid activations to compare how each activation constructs a latent space. 

The 2048-point first-order difference input augmentation is replaced with the 12-point one-hot encoded chroma feature explained above. Furthermore, in this work three separate topologies are explored by varying the bottleneck layer's width -- one with two neurons, one with three neurons, and one with eight neurons.  

\begin{figure}[ht]
\centerline{\includegraphics[width=0.4\paperwidth]{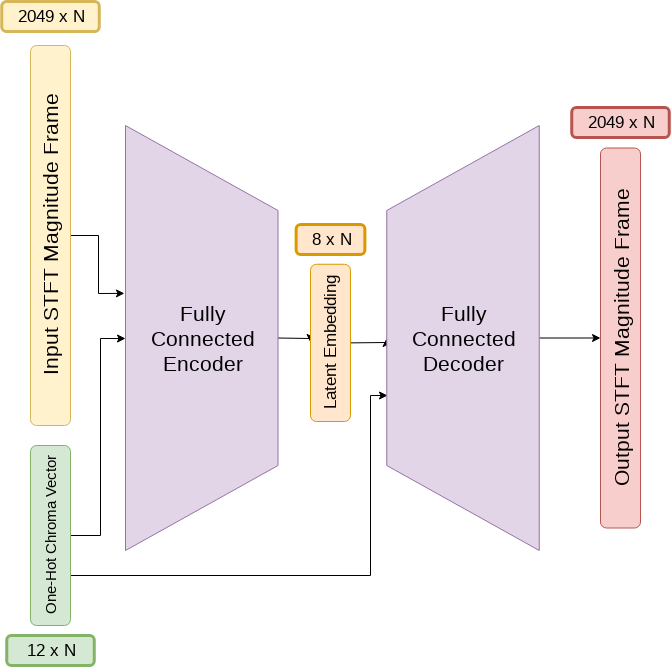}}
\caption{Network diagram with eight neuron bottleneck and chroma skip connection}
\label{diagram}
\end{figure}

Residual and skip connections are used in autoencoder design to help improve performance as network depth increases \cite{residual}. In this work, the 12-point one-hot encoded chroma feature input augmentation is passed directly to the autoencoder's bottleneck layer. Models with and without this skip connection are trained to compare how the skip connection affects the autoencoder's latent space. Figure \ref{diagram} depicts our architecture with the chroma skip connection and eight neuron latent space. Note that for our two neuron model, the $8 \times N$ latent embedding would become a $2 \times N$ latent embedding, and similarly would become a $3 \times N$ for our three neuron model.

\section{Corpora}
In this work a multi-layer neural network autoencoder is trained to learn representations of musical timbre. The aim is to train the autoencoder to contain high level descriptive features in a low dimensional latent space that can be easily manipulated by a musician. As in the formulation above, dimension reduction is imposed at each layer of the encoder until the desired dimensionality is reached. All audio used to generate the corpora for this work is stored as a 16-bit PCM wav file with 44.1kHz sampling rate. 

The various corpora used to train the autoencoding neural network are formed by taking 2049 points from a 4096-point magnitude STFT $s_{n}(m)$ as its target, where $n$ denotes the frame index of the STFT and $m$ denotes the frequency index, with 75\% frame overlap. The Hann window is used in all cases. Each frame is normalized to $[0,1]$. This normalization tasks the autoencoder to encode solely the timbre of an input observation and ignore its loudness relative to other observations within the corpus. These corpora were not mixed for training; models were only trained on each corpus separately.

\subsection{MicroKORG Dataset}

Two corpora were created by recording C Major scales from a MicroKORG synthesizer/vocoder. In both cases, 70 patches make up the training set, 5 patches make up the validation set, and 5 patches make up the test set. These patches ensured that different timbres are present in the corpus. To ensure the integrity of the testing and validation sets, the dataset is split on the ``clip'' level. This means that the frames in each of the three sets are generated from distinct passages in the recording, which prevents duplicate or nearly duplicate frames from appearing across the three sets.  

The first corpus is comprised of approximately $79,000$ magnitude STFT frames, with an additional $6,000$ frames held out for validation and another $6,000$ for testing. This makes the corpus  $91,000$  frames in total, or roughly $35$ minutes of audio. The audio used to generate these frames is composed of five octave C Major scales recorded from a MicroKORG synthesizer/vocoder across 80 patches. 

The second corpus is a subset of the first. It is comprised of one octave C Major scales starting from concert C. Approximately $17,000$ frames make up the training set, with an additional $1,000$ frames held out for validation and another $1,000$ for testing. This makes the subset  $19,000$  frames in total, or roughly $7$ minutes of audio.

By restricting the corpus to single notes played on a MicroKORG, the autoencoder needs only to learn higher level features of harmonic synthesizer content. These tones often have time variant timbres and effects, such as echo and overdrive. Thus the autoencoder is also tasked with learning high level representations of these effects. 

\subsection{TU-Note Violin Sample Library}
A third corpus was created using a portion of the TU-Note Violin Sample Library \cite{tu-note}. The dataset consists of recordings of a violin in an anechoic chamber playing single sounds, two-note sequences, and solo performances such as scales and compositions. The single notes were used to construct a training corpus, and the solo performances were cut into two parts to form the validation and test sets. These two parts were split on the ``clip'' level to ensure that no frames from the same passages were found across the validation and test sets. Approximately $91,000$ frames make up the training set, with an additional $10,000$ frames held out for validation and another $10,000$ for testing. This makes the subset  $111,000$  frames in total, or roughly $43$ minutes of audio. Here, the autoencoder is tasked with learning the difference in timbre one can here when a violin is played at different dynamic levels, semitones, and with different stroke techniques.

\subsection{Training Setup}
All models were trained for 300 epochs using the ADAM method for stochastic optimization \cite{ADAM}, initialized with a learning rate of $5 \times 10^{-4}$. Mean squared error was used as the cost function, with an L2 penalty of $10^{-7}$ \cite{L2}. All training utilized one NVIDIA Quadro P2000 GPU, and all networks were implemented using Keras 2.2.4 \cite{keras} with Tensorflow-GPU 1.9.0 \cite{tensorflow} as a backend. 

\section{Results}

\begin{table}[t] 
  \centering
  \begin{tabular}{|c|c|c|c|}
    \hline \textbf{Input Augmentation}     	& \textbf{Test Set MSE}	& \textbf{Training Time}     \\\hline
    No Append	  & $3.185 \times 10^{-4} $  &	35 minutes \\	
    $1^{st}$ Order Diff  & $3.001 \times 10^{-4} $  &	44 minutes	 \\
    One-Hot Chroma    & $2.920 \times 10^{-4} $  &	37 minutes	 \\\hline
  \end{tabular} 
  \caption{Five Octave Dataset Autoencoder holdout set MSE loss and training time}
  \label{5_octave_loss}
\end{table}

\begin{table}[t] 
  \centering
  \begin{tabular}{|c|c|c|c|}
    \hline \textbf{Bottleneck Activation}	& \textbf{Skip?}	& \textbf{Test Set MSE}      \\\hline
    LReLU	  & No  & $3.516 \times 10^{-4} $   \\	
    LReLU  & Yes   & $3.598 \times 10^{-4} $  	 \\
    Sigmoid  & No  & $3.358 \times 10^{-4} $  	 \\
    Sigmoid    & Yes  & $3.472 \times 10^{-4} $  	 \\\hline
  \end{tabular} 
  \caption{One Octave Dataset Autoencoder holdout set MSE loss, 2 neuron bottleneck}
  \label{1_octave_loss}
\end{table}

\begin{table}[t] 
  \centering
  \begin{tabular}{|c|c|c|c|}
    \hline \textbf{Bottleneck Activation}	& \textbf{Skip?}	& \textbf{Test Set MSE}  & \textbf{Training Time}    \\\hline
    LReLU	  & No  & $7.731 \times 10^{-4} $  &	41 minutes \\	
    LReLU  & Yes   & $6.478 \times 10^{-4} $  &	58 minutes	 \\
    Sigmoid  & No  & $8.900 \times 10^{-4} $  &	43 minutes	 \\
    Sigmoid    & Yes  & $6.388 \times 10^{-4} $  &	43 minutes	 \\\hline
  \end{tabular} 
  \caption{TU-Note Violin Sample Library Dataset Autoencoder holdout set MSE loss and training time, two neuron bottleneck}
  \label{TU-Note_loss}
\end{table}

\begin{table}[t] 
  \centering
  \begin{tabular}{|c|c|c|c|}
    \hline  \textbf{Corpus}&\textbf{Skip?}	& \textbf{Test Set MSE} & \textbf{Training Time}     \\\hline
     One Octave & No  & $3.228 \times 10^{-4} $  & 	8 minutes \\
     One Octave & Yes  & $3.055 \times 10^{-4} $  & 8 minutes	 \\
     Violin & No  & $8.545 \times 10^{-4} $  & 44 minutes	 \\
     Violin & Yes  & $5.763 \times 10^{-4} $  & 45 minutes	 \\\hline
  \end{tabular} 
  \caption{Sigmoid Bottleneck Autoencoder holdout set MSE loss and training time, three neuron bottleneck}
  \label{1_octave_loss_3D}
\end{table}

Table \ref{5_octave_loss} shows the performance of three autoencoders with an eight neuron bottleneck layer using LReLU activations trained on the five octave MicroKORG corpus. The model with the chroma augmentation outperforms both the first-order difference augmentation and no augmentation models. Moreover, the chroma augmentation only increases training time by two minutes. Therefore, the rest of the models in this work utilize the chroma input augmentation. 

Table \ref{1_octave_loss} show the performance of four autoencoders with a two neuron bottleneck layer trained on the one octave MicroKORG corpus. Models used either the LReLU or sigmoid activation for the bottleneck, and either did or did not utilize a chroma skip connection. All models took eight minutes to train. Both sigmoid models outperformed each LReLU model, and the sigmoid model with no skip connection performed the best.

Table \ref{TU-Note_loss} show the performance of four autoencoders with a two neuron bottleneck layer trained on the TU-Note Violin Sample Library corpus. Models used either the LReLU or sigmoid activation for the bottleneck, and either did or did not utilize a chroma skip connection. With this corpus, the chroma skip connection significantly improved the reconstruction error for both sigmoid and LReLU activations. Furthermore, the sigmoid activation with the chroma skip connection outperformed all models.

With these results in mind, two models were trained on the one octave MicroKORG corpus using a three neuron bottleneck with sigmoid activations: one with the chroma skip connection, and one without. Two more models with corresponding topologies were trained on the TU-Note Violin Sample Library corpus. Table \ref{1_octave_loss_3D} shows the reconstruction performance of each model. In this case, the models with the chroma skip connection outperformed the models without. 

\subsection{Latent Embeddings}
When designing an autoencoder for musicians to use in timbre synthesis, it is important not only to measure the network's reconstruction error, but also to characterize the latent space's embedding. The software synthesizer implemented in \cite{colonel2} allows a musician to chose a point in the autoencoder's latent space and generate its corresponding timbre. By exploring the latent space, the musician can explore an embedding of timbre.

A clear understanding of the boundedness of an embedding ensures that a musician can fully explore the latent space of an arbitrary training corpus, and a clear understanding of the density of the latent embedding can help a musician avoid portions of the latent space that will generate unrealistic examples while interpolating between two encoded timbres \cite{manifold} \cite{sampling}. 

Recent work has attempted to encourage an autoencoder to interpolate in a ``semantically continuous'' manner \cite{interpolate}. The authors sample from their autoencoder's latent space along a line that connects two points to demonstrate this meaningful interpolation. The authors also characterize their latent space using a method proposed by \cite{unsupervised}, where an unsupervised clustering accuracy is measured to see how well ground truth labels are separated in the latent space.  In the case of our work, however, we are less concerned with how clusters separate in the latent space and more concerned with how uniform samplings of the latent space produce note classes and timbres.    

\begin{center}
\begin{figure}[ht]
\centerline{\includegraphics[width=0.4\paperwidth]{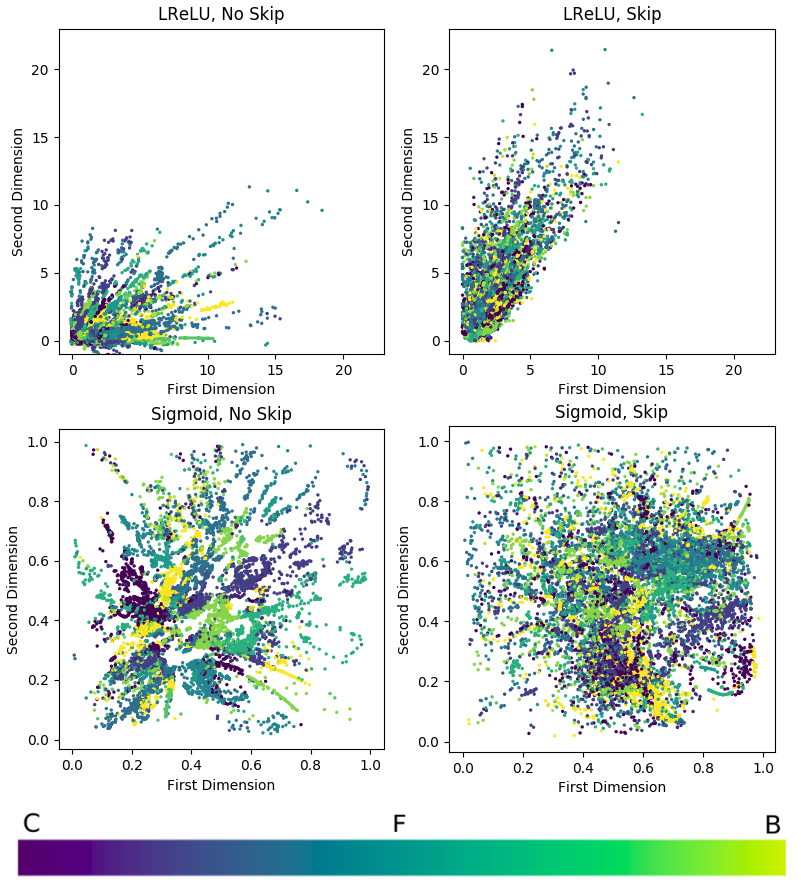}}
\caption{2D embeddings of the One Octave MicroKORG Corpus}
\label{2D_OO}
\end{figure}
\end{center}

\begin{figure}[ht]
\centerline{\includegraphics[width=0.39\paperwidth]{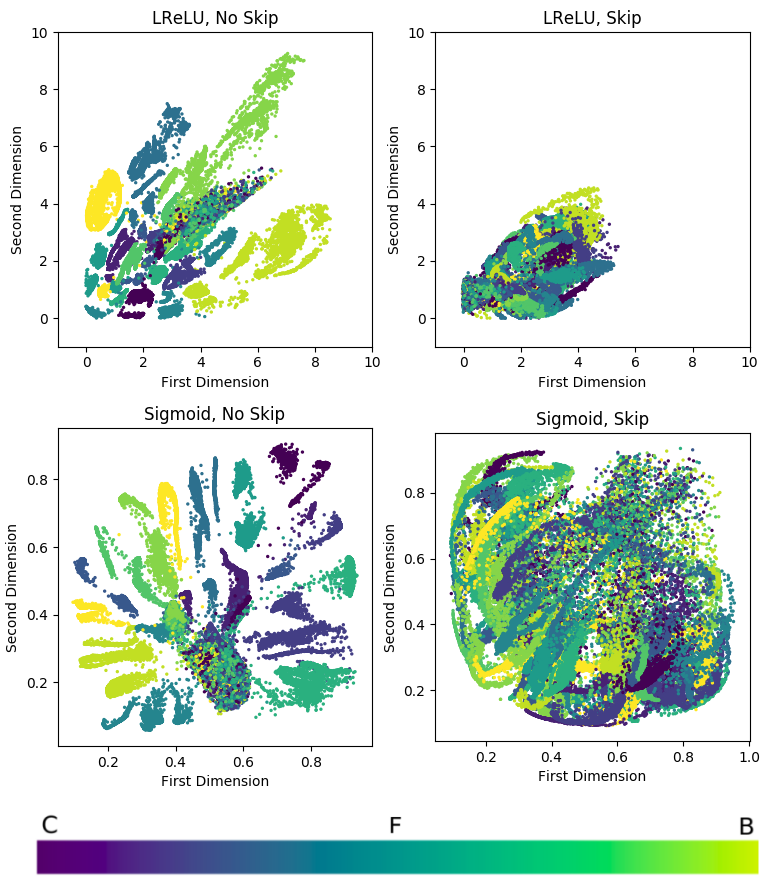}}
\caption{2D embeddings of the TU-Note Violin Sample Library Corpus}
\label{2D_VIO}
\end{figure}

We begin with a visual inspection of the training set embeddings produced by the eight distinct autoencoders referred to in Tables \ref{1_octave_loss} and \ref{TU-Note_loss}. Figure \ref{2D_OO} shows the embeddings for the one octave MicroKORG corpus, and Figure \ref{2D_VIO} shows the embeddings for the TU-Note Violin Sample Library corpus. Models trained with the LReLU bottleneck activation are plotted in the top row, and models trained with the sigmoid bottleneck activation are plotted in the bottom row. Models trained without the chroma skip connection are plotted in the left column, and models trained with the chroma skip connection are plotted in the right column. Each note class is plotted as one color (i.e. C is dark blue, F is teal, B is yellow) using a perceptually uniform colormap.

In all cases, the chroma skip connection appears to encourage the embedding to be denser and contain fewer striations. Note that by definition, all models with sigmoid activations are bounded by $(0,1)$. On the other hand, the models with LReLU activation vary their bounds greatly.  Moreover, the first and second dimensions of the LReLU embeddings appear to have linear correlations, rather than populating the latent space in a more uniform manner. As such we move forward using only sigmoid activations at the bottleneck.

\begin{figure*}[ht]
\centerline{\includegraphics[width=0.63\paperwidth]{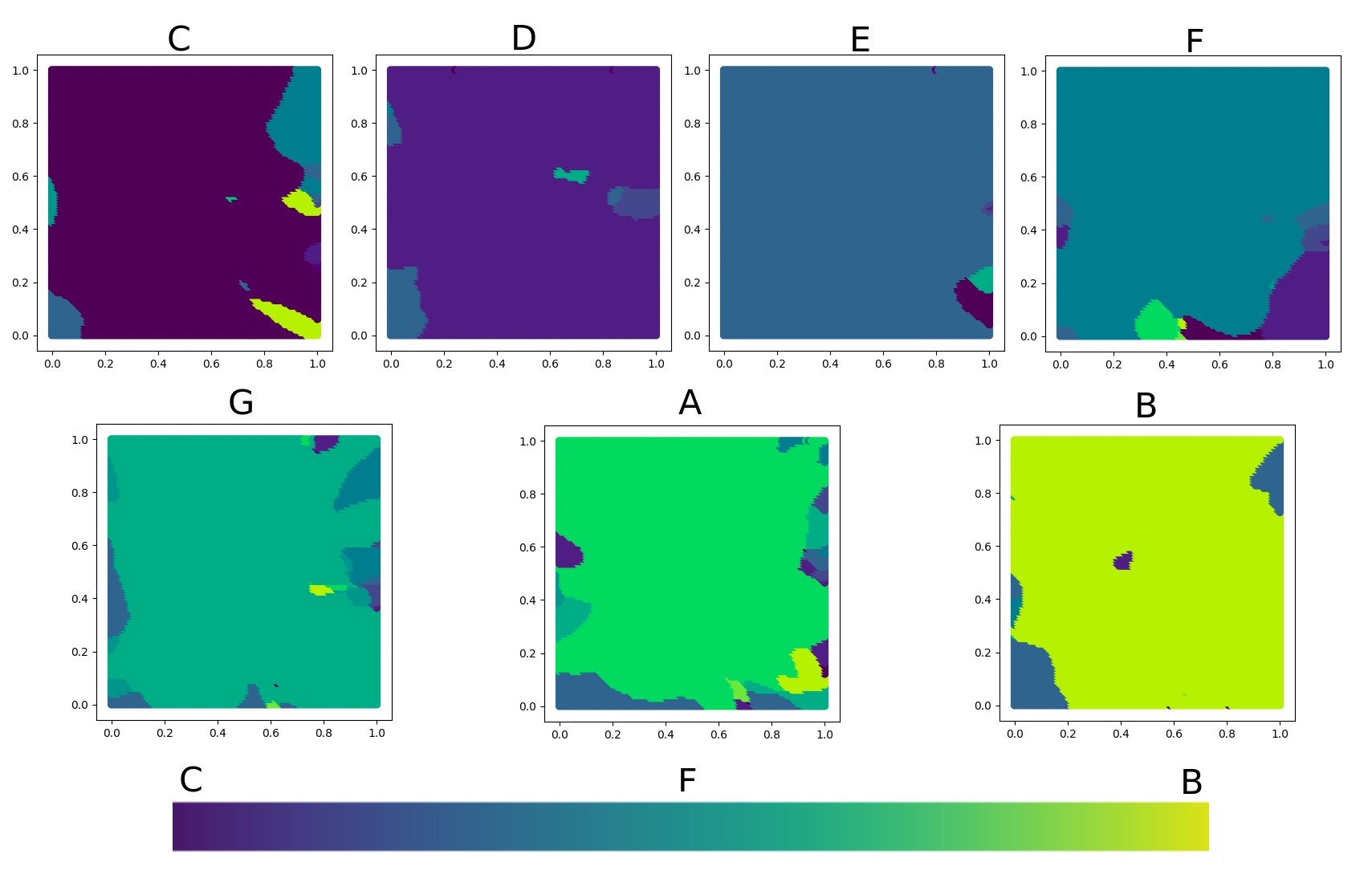}}
\caption{2D embeddings of the One Octave Corpus}
\label{oo_notes}
\end{figure*}

\begin{figure*}[ht]
\centerline{\includegraphics[width=0.63\paperwidth]{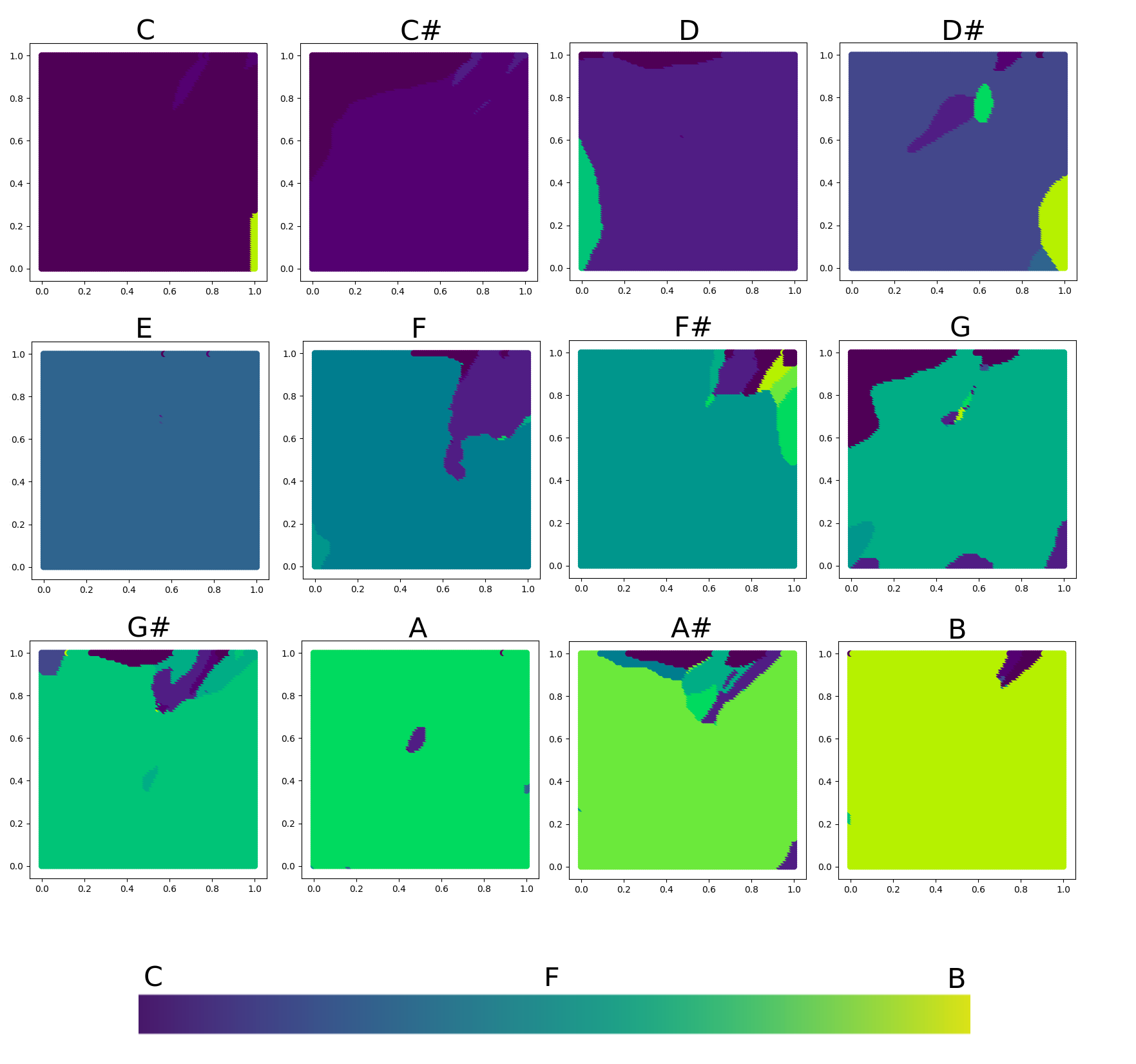}}
\caption{2D embeddings of the TU-Note Violin Sample Library Corpus}
\label{vio_notes}
\end{figure*}

A full accounting of the two neuron sigmoid bottleneck autoencoder's latent space is shown in Figures \ref{oo_notes} and \ref{vio_notes}. These graphs were created by setting the chroma conditioning vector to a given note class, and then sampling the autoencoder's latent space using a $350 \times 350$ point mesh grid. Each note class is plotted as one color (i.e. C is dark blue, F is teal, B is yellow) using a perceptually uniform colormap. We observe that the autoencoder is able to use the majority of the alloted two dimensions to produce timbres that match the conditioned chroma vector. We note that most mismatches occur near the boundaries of the latent space. We suspect this may be caused by the asymptotic behavior of the sigmoid function coupled with the L2 penalty encouraging the network to choose smaller weights, though a full characterization is outside the scope of this paper. 

This mesh sampling procedure was repeated for the three neuron and eight neuron sigmoid bottleneck models. Due to compuational constraints, the three neuron model used a $50 \times 50 \times 50$ mesh and the eight neuron model used a $5 \times 5 \times ... \times 5$ mesh. The accuracies of the model samplings are shown in Table \ref{Samplings}. We suspect that some of the decreases in prediction accuracy as the number of neurons in the bottleneck increases may be due in part to the coarser meshes over-weighing samplings near the boundaries of the latent space, though a full characterization is outside the scope of this paper. 

\begin{table*}[t] 
  \centering
  \begin{tabular}{|c|c|c|c|c|c|c|c|c|c|c|c|c|c|c|c|}
    \hline \textbf{Model} & \textbf{Mesh Length} &\textbf{C}& \textbf{C\#}& \textbf{D}& \textbf{D\#}& \textbf{E}& \textbf{F}& \textbf{F\#}& \textbf{G}& \textbf{G\#}& \textbf{A}& \textbf{A\#}& \textbf{B}    \\\hline
     2D One Octave & 350 &$.896$& --& $.941$& --& $.981$& $.884$& --& $.908$& --& $.863$& --& $.937$ \\
     3D One Octave & 50 &$.827$& --& $.857$& --& $.847$& $.953$& --& $.851$& --& $.904$& --& $.933$   	 \\
     8D One Octave & 5 &$.741$& --& $.650$& --& $.949$& $.526$& --& $.605$& --& $.834$& --& $.660$   	 \\
     2D Violin & 350 &$.980$& $.840$& $.941$& $.902$& $.999$& $.846$& $.905$& $.864$& $.892$& $.992$& $.884$& $.984$ 	 \\
     3D Violin & 50 &$.942$& $.629$& $.999$& $.692$& $.999$& $.997$& $.659$& $.980$& $.999$& $.999$& $.885$& $.977$ 	 \\
     8D Violin & 5 &$.837$& $.833$& $.950$& $.844$& $.814$& $.805$& $.956$& $.782$& $.820$& $.844$& $.805$& $.920$ 	 \\\hline
  \end{tabular} 
  \caption{Percent of sampled outputs matching conditioned chroma skip vector}
  \label{Samplings}
\end{table*}

\section{Python Implementation}
As outlined in \cite{colonel2}, a spectrogram with no phase information can be generated via bypassing the network's encoder and passing latent activations to the decoder. To generate the true phase of this spectrogram, the real-time phase gradient heap integration algorithm can be used \cite{phase}. However, to decrease the computational overhead involved in our algorithm, we store the stripped the phase of a white noise audio signal and use it to invert the generated spectrogram. 

Our implementation is purely Python, using Tkinter as our GUI backend. Once a user selects a trained decoder to sample from, Keras loads the model into memory. The user is presented with sliders that correspond to each neuron in the model's bottleneck, and a twelve-value radio button is used to set the chroma conditioning vector. The Pyaudio library provides Python bindings to PortAudio \cite{portaudio}, and handles the audio stream output. 

Our implementation has been made available at \textit{github.com/JTColonel/manne}, along with code to create a corpus from an audio file for training, code to train a model, and code to plot the samplings of a model's latent space. We have tested our implementation on a laptop with an Intel Core i7-8750H CPU @ 2.20GHz × 12 with 16GB of RAM.

We also provide code to train and sample from Variational Autoencoder implementation (specifically a $\beta$-VAE \cite{beta}), with a word of caution. We found that all of our trained models exhibited posterior collapse \cite{partial_collapse}, wherein the variational distribution would closely match the uninformative prior for a subset of latent variables, and the rest of the latent variables would output high mean, low variance samplings. Moreover, we did not find that the non-conditioned $\beta$-VAE disentangled the note class from timbre. We found that the note class would change when varying any one latent dimension while fixing the rest.  Unfortunately a full treatment of this behavior is outside the scope of this paper. 

\section{Conclusion}
We present an improved neural network autoencoder topology and training regime for use in timbre synthesis and interpolation. By using a one-hot encoded chroma vector as both an augmentation to the autoencoder's input and a conditioning vector at the autoencoder's bottleneck, we improve the autoencoder's reconstruction error at little additional computational cost. Moreover, we characterize how this conditioning vector shapes the autoencoder's usage of its latent space. We provide an open source implementation of our architecture in Python, which can sample from its latent space in real-time without the need for powerful computing hardware.

\bibliographystyle{IEEEtranS.bst}
\bibliography{IEEEabrv,bibliography.bib}

\end{document}